\documentclass[11pt]{article}
\usepackage{moriond}

\def\be{\begin{equation}}
\def\ee{\end{equation}}
\def\bea{\begin{eqnarray}}
\def\eea{\end{eqnarray}}

\long\def\comment#1{ }

\newcommand{\beq}{\begin{eqnarray}}
\newcommand{\eeq}{\end{eqnarray}}

\newcommand{\eqn}[1]{Eq.~(\ref{#1})}

\newcommand{\rmd}{{\rm d}}
\newcommand{\rme}{{\rm e}}

\newcommand{\order}[1]{\mcal{O}{(#1)}}
\newcommand{\mcal}{\mathcal}
\newcommand{\abar}{\bar{\alpha}}

\def\lesssim{\mathrel{
   \rlap{\raise 0.511ex \hbox{$<$}}{\lower 0.511ex \hbox{$\sim$}}}}
   \def\gtrsim{\mathrel{%
   \rlap{\raise 0.511ex \hbox{$>$}}{\lower 0.511ex \hbox{$\sim$}}}}

\newcommand{\Photo}{}

\begin{document}
\vspace*{4cm}
\title{FROM JET QUENCHING TO WAVE TURBULENCE}

\author{E. IANCU}

\address{Institut de Physique Th\'eorique de Saclay\\
 F-91191 Gif-sur-Yvette, FRANCE}

\maketitle\abstracts{We discuss average properties of the gluon
cascade generated by an energetic parton propagating through a dense QCD medium.
The cascade is mostly made with relatively soft gluons, whose production is not
suppressed by the LPM effect. Unlike for usual QCD cascades in the vacuum, where
the typical splittings are very asymmetric (soft and collinear), the medium--induced 
branchings are quasi--democratic and lead to wave turbulence. 
This results in a very efficient mechanism for the transport of  energy at large angles
with respect to the jet axis, which might explain the di--jet asymmetry observed in Pb--Pb 
collisions at the LHC. 
}

One remarkable phenomenon discovered in the lead--lead collisions at the LHC is the 
{\em di--jet asymmetry}, a strong imbalance between the energies of two back--to--back jets. 
This asymmetry is commonly attributed to the effect of the interactions of one of the two jets with the 
dense QCD matter that it traverses, while the other leaves the system unaffected. Originally identified 
\cite{Aad:2010bu,Chatrchyan:2011sx} as missing energy, this phenomenon has been subsequently 
shown \cite{Chatrchyan:2012ni} to consist in the transport of a sizable part (about $10\%$) 
of the total jet energy by {\em soft particles} towards {\em large angles}. 
The total amount of energy which is thus lost by the jet, $\sim 20$~GeV, 
is considerably larger than the typical transverse momentum,  $\sim 1$~GeV, of a parton in  the medium.
In that sense, the effect is large and potentially non--perturbative.
Yet, there exists a mechanism within perturbative QCD which can
naturally explain the energy loss at large angles: the BDMPSZ  mechanism for 
medium--induced gluon radiation (from Baier, Dokshitzer, Mueller, Peign\'e, Schiff 
\cite{Baier:1996kr} and Zakharov \cite{Zakharov:1996fv}). Most previous studies within this approach 
have focused on the energy lost by the leading particle, which is controlled by rare and relatively hard
emissions at small angles. More recently, in the wake of the LHC data, the attention has been shifted 
towards softer emissions, for which the effects of {\em multiple branching} become important.
The generalization of the BDMPSZ formalism to multiple branchings has only recently been given
\cite{CasalderreySolana:2011rz,Blaizot:2012fh}, and this turns out to have interesting 
physical consequences \cite{Blaizot:2013hx}.
These recent developments will be briefly reviewed in what follows.

The BDMPSZ mechanism relates the radiative energy loss
by an energetic parton propagating through a dense QCD medium (`quark--gluon plasma') to the
transverse momentum broadening via scattering off the medium constituents. A central
concept is the {\em formation time} $\tau_f(\omega)$ --- the typical times it takes 
a gluon with energy $\omega \ll E$ to be emitted.
($E$ is the energy of the original parton, a.k.a. the `leading particle'.) The gluon starts as a virtual 
fluctuation which moves away from its parent parton via quantum diffusion:
the transverse \footnote{The `transverse directions' refer to the 2--dimensional plane orthogonal 
to the 3--momentum of the leading particle, which defines the `longitudinal axis'.}
separation $b_\perp$ grows with time as $b_\perp^2 \sim \Delta t/\omega$.
The gluon can be considered as `formed' when it loses coherence w.r.t to its source, meaning that  
$b_\perp$ is at least as large as the gluon transverse wavelength $\lambda_\perp
= 1/k_\perp$.  But the gluon transverse momentum $k_\perp$ is itself increasing with time, 
via collisions  which add random kicks $\Delta k_\perp$
at a rate given by the {\em jet quenching parameter} $\hat q$ :
$\Delta k_\perp^2\sim \hat q \Delta t$. The `formation' condition, $b_\perp\gtrsim 1/\Delta k_\perp$ for
$\Delta t\gtrsim\tau_f$, implies (below, $k_f$ and $\theta_f$ are the typical values of the
gluon transverse momentum and emission angle at the time of formation)
\beq\label{thetaf} 
\tau_f(\omega)\,\simeq \,\sqrt{\frac{2\omega}{\hat q}}\,, \qquad  k_{f}^2 \,=\,\hat q \tau_f(\omega)
\,\simeq\,
 (2\omega\hat q)^{1/2}\,,\qquad \theta_f\,\simeq\,\frac{k_{f}}{\omega}\,\simeq\,
 \left(\frac{2 \hat q}{\omega^3}\right)^{1/4}.\eeq
\eqn{thetaf} applies as long as $\ell\ll \tau_f(\omega) < L$, where $L$ is the length of the medium
and $\ell$ is the mean free path between successive collisions.
The second inequality implies an upper limit on the energy 
of a gluon that can be emitted via this mechanism, and hence a lower limit on
the emission angle: $\omega \lesssim \omega_c\equiv \hat q L^2/2$ and 
$\theta_f\gtrsim  \theta_c\equiv 2/(\hat q L^3)^{1/2}$. The BDMPSZ regime
corresponds to \footnote{One can estimate $\hat q\simeq m_D^2/\ell$, where $m_D$ is the Debye screening 
mass in the medium. \eqn{thetaf} applies to a medium whose size $L$ is much larger
than both the screening length $1/m_D$ and the mean free path $\ell$.} 
$\hat q L^3\gg 1$ and hence $\theta_c\ll 1$. Choosing $\hat q=1\,{\rm GeV}^2/{\rm fm}$ 
(the weak coupling estimate \cite{Baier:1996kr}  for a QGP with temperature 
$T=250$~MeV) and $L= 4$~fm, one finds
$ \omega_c\simeq 40$~GeV and $\theta_c\simeq 0.05$.

\eqn{thetaf} shows that the relatively soft gluons with $\omega\ll \omega_c$ have \texttt{(i)} 
short formation times $\tau_f(\omega)\ll L$ and \texttt{(ii)} large formation angles $\theta_f\gg\theta_c$.
Both properties are important for us here. Property \texttt{(i)} implies that such gluons are 
produced {\em abundantly}~: their emission can be initiated at any place inside the medium, hence the
associated spectrum (below, $\bar\alpha\equiv \alpha_s N_c/\pi$)
  \beq\label{spec}
\omega \frac{\rmd N}{\rmd \omega}
 \,\simeq\,\frac{\alpha_s N_c}{\pi}\,\frac{L}{\tau_f(\omega)}\,=\,
 \bar\alpha\sqrt{\frac{\omega_c}{\omega}}\,, \eeq
is enhanced by a factor  ${L}/{\tau_f(\omega)}\gg1$, which expresses the relative longitudinal
phase--space available for their emission. Property \texttt{(ii)} shows that the soft gluons 
have the potential to transport a part of the jet energy towards large angles. This looks like a
small effect, since soft gluons carry only little energy, but this is
enhanced by multiple emissions, as we shall see.

\eqn{spec} is the BDMPSZ spectrum for a soft, 
medium--induced, gluon emission \cite{Baier:1996kr,Zakharov:1996fv}. 
Note that the emission probability is small,  of $\order{\abar}$, for a
relatively hard gluon with $\omega\sim \omega_c$. This is a consequence of the 
{\em LPM  effect} (from Landau, Pomeranchuk, Migdal) --- the fact that one needs
a large number $\tau_f(\omega)/\ell\gg 1$ of successive scatterings in order to produce
a single gluon. Such a rare but hard emission dominates the 
average energy loss by the leading particle, $\Delta E\equiv \int^{\omega_c}
\rmd\omega \, \omega ({\rmd N}/{\rmd \omega})\sim\abar\omega_c$\,, but it cannot
contribute to the observed di--jet asymmetry, because a hard gluon 
propagates at a very small angle $\sim \theta_c$ w.r.t. the jet axis.

\begin{figure}[t]
	\centerline{\includegraphics[width=0.7\linewidth]{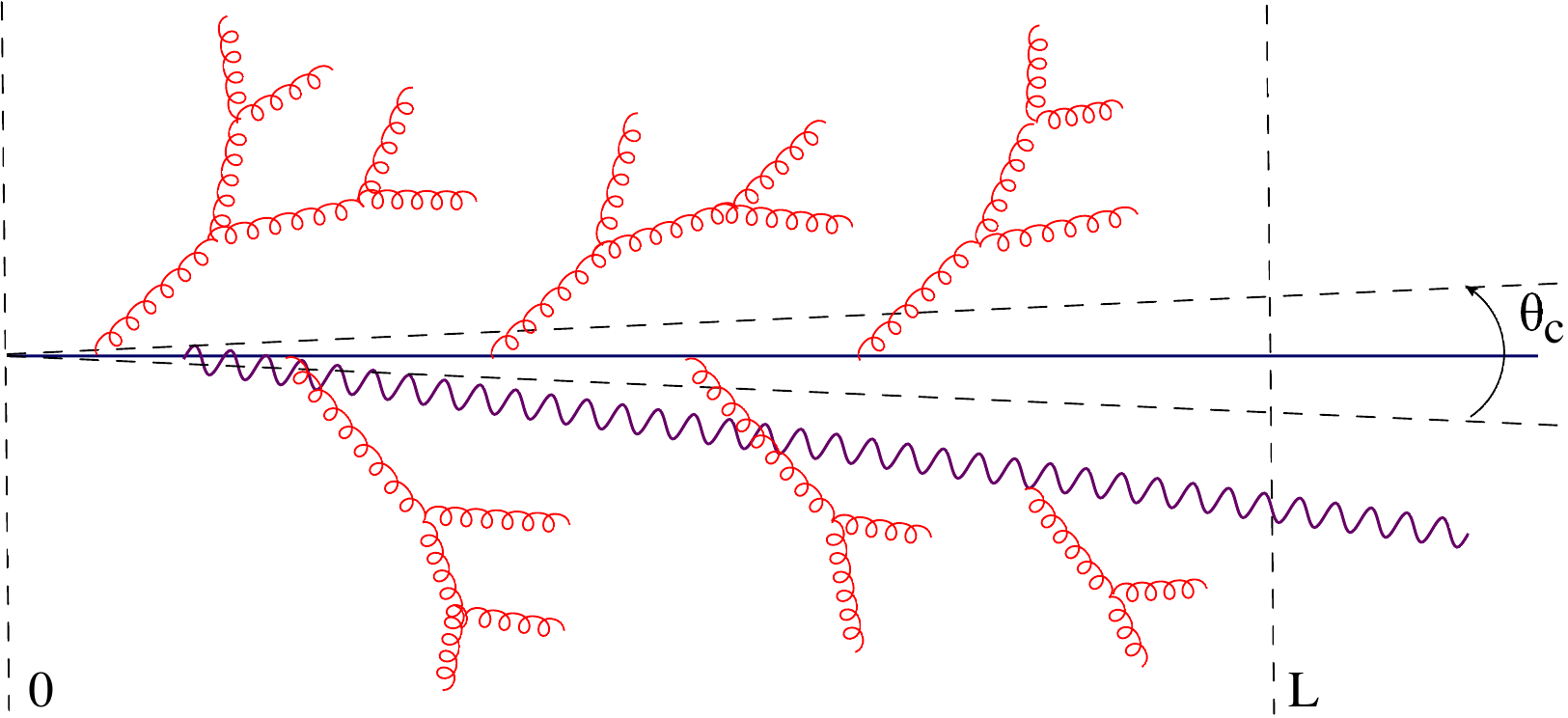}}
	\caption{The jet produced in a  `rare event' which, besides the leading particle
	and the (quasi--deterministic) soft gluon cascades at large angles, also
contains a hard gluon with $\omega \sim\omega_c$, which propagates at
a small angle $\theta\sim \theta_c$.}
\label{fig:jet}
\end{figure}

To understand di--jet asymmetry,
one rather needs to focus on soft gluon emissions at large angles. For such emissions,
the effects of {\em multiple branching} become important, as we now explain. 
The probability for a single emission, as measured by the spectrum (\ref{spec}), exceeds unity when
$\omega\lesssim \abar^2\omega_c$.  In this non--perturbative regime at 
small $\omega$, \eqn{spec} must be corrected
to account for multiple emissions by the leading particle, and also for the subsequent branching 
of the soft primary gluons into even softer ones (thus leading to gluon cascades;
see Fig.~\ref{fig:jet}).  The soft branchings are quasi--deterministic and can be observed in
an event--by--event basis.

\comment{Besides the leading particle, the `jet'  in Fig.~\ref{fig:jet} involves a relatively hard 
gluon ($\omega\sim\omega_c$) emitted at a small angle $\theta\sim \theta_c$
w.r.t. the jet axis, accompanied by a `rain' of soft gluons, which propagate at large angles
and further branch into gluon cascades.}

The treatment of multiple branching is {\em a priori} complicated by interference effects 
between emissions from different partonic sources. For the case of a jet evolving into the vacuum,
such effects are known to lead to {\em angular ordering} between the successive emissions.
Interference effects for medium--induced gluon radiation started to be investigated only recently
\cite{CasalderreySolana:2011rz,MehtarTani:2011tz,MehtarTani:2012cy,Blaizot:2012fh}. The respective
analysis is quite involved, but its main conclusion is very simple  \cite{Blaizot:2012fh}: 
the interference effects for soft emissions of the BDMPSZ  type are 
{\em negligible}, since suppressed by a factor $\tau_f(\omega)/ L\ll 1$.  
To understand this result, recall that, in order to  interfere with each other, two
emitters must be {\em coherent} which each other, a situation which can occur 
if they have a common ancestor. However, the partons produced by a medium--induced branching
lose their mutual {\em colour coherence} already during the formation process, because they randomly
scatter in the medium. Accordingly, they can interfere with each other only during 
a short time $\tau_f(\omega)$, which gives a small phase--space whenever $\omega\ll \omega_c$.
This implies that successive medium--induced emissions can be
considered as {\em independent} of each other and taken into account via a {\em probabilistic 
branching process},  in which the BDMPSZ spectrum plays the role of
the elementary branching rate \footnote{Such a classical branching process, obtained by iterating the 
single BDMPSZ emission, has already been used in applications to phenomenology, albeit on
a heuristic basis \cite{Baier:2000sb,Baier:2001yt,Jeon:2003gi}.}.

The general branching process is a Markovian process in 3+1 dimensions which describes the
gluon distribution in energy ($\omega$) and transverse momentum ($k_\perp$), and
its evolution when increasing the medium size $L$. This process is well
suited for numerical studies via Monte--Carlo simulations. But analytic results have also been
obtained \cite{Blaizot:2013hx}, for a simplified process in 1+1 dimensions, 
which describes the energy distribution alone. 
These results lead to an interesting physical picture that we shall now describe.

\begin{figure}[t]
	\centerline{\includegraphics[width=0.9\linewidth]{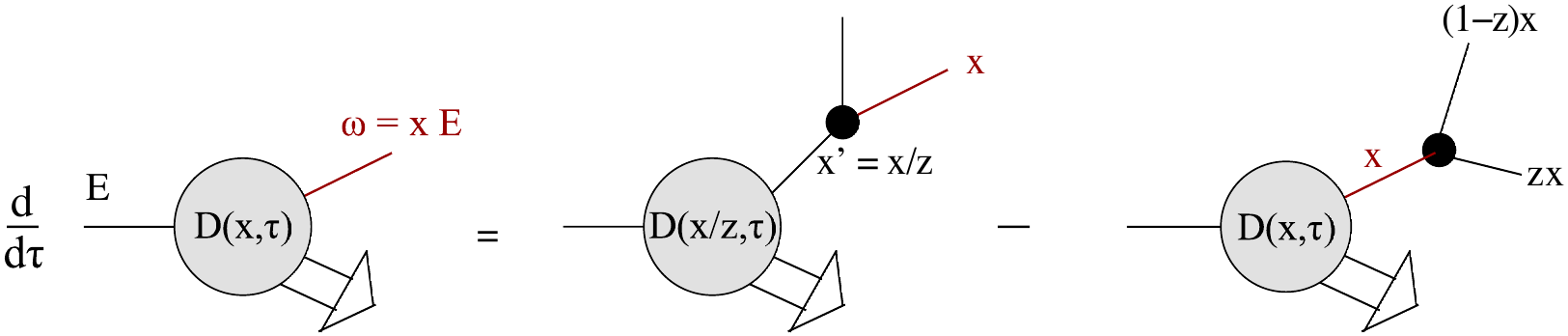}}
	\caption{The change in  the gluon spectrum $D(x,\tau)\equiv x({\rmd N}/{\rmd x})$
	due to one additional branching $g\to gg$.}
\label{fig:rate}
\end{figure}


To that aim, it is convenient to focus on the {\em gluon spectrum} 
$D(x,\tau)\equiv x({\rmd N}/{\rmd x})$, where $x\equiv \omega/E$ is the energy fraction carried by 
a gluon from the jet and the `evolution time' $\tau$ is the medium size in 
dimensionless units, as defined in  \eqn{Ds} below. 
The quantity $D(x,\tau)$ obeys an evolution equation \cite{Blaizot:2013hx,Baier:2000sb,Jeon:2003gi},
which is formally similar to the DGLAP equation describing the
fragmentation of a jet in the vacuum: a rate equation, which involves a `gain' term 
and a `loss' term and is illustrated in Fig.~\ref{fig:rate}. 
The `gain' term describes the increase in the number of gluons with a given $x$
via radiation from gluons with a larger $x'=x/z$, with any $x<z<1$. The `loss' term expresses the 
decrease in the number of gluons at $x$ via their decay $x\to zx, (1-z)x$, with any $0<z<1$.
What is specific to the problem at hand, is the particular form of the  
in--medium branching rate --- the BDMPSZ spectrum in \eqn{spec} ---, which is very
different from the DGLAP splitting function.
This difference has important physical consequences, that can be best appreciated by comparing 
the respective solution $D(x,\tau)$ to the BDMPSZ spectrum
for a single emission and to the DGLAP spectrum for gluon evolution in the vacuum.

By construction, the BDMPSZ spectrum coincides with the first iteration to the evolution equation. For
soft gluons ($x\ll 1$), it is given by \eqn{spec}, which in our new notations reads
 \beq\label{Ds}
 D^{(1)}(x\ll 1,\tau)\,\simeq\,\frac{\tau}{\sqrt{x}}\,,\qquad{\rm with}\quad
 \tau\,\equiv\,\bar \alpha\sqrt{\frac
 {\hat q}{E}}\,L\,.
 \eeq
This spectrum increases quite fast with $1/x$ and thus predicts that a non--negligible
fraction of the total radiated energy is emitted directly at large angles: the energy fraction
transported via a single gluon emission at angles larger than a given value $\theta_0$ is estimated as
 \beq\label{E1}
  {\cal E}^{(1)}(\theta > \theta_0,\tau)=\int_0^{x_0} \rmd x D^{(1)}(x,\tau)
  \,\simeq\,2\tau \sqrt{x_0}\quad\mbox{with}\quad 
   \theta_0\,\simeq\,
 \left(\frac{2 \hat q}{x_0^3 E^3}\right)^{1/4}\,.
  \eeq
This quantity rapidly decreases with increasing $\theta_0$ (i.e. with decreasing $x_0$), 
showing that direct radiation by the leading particle cannot explain 
the large energy loss at large angles observed in relation with the di--jet asymmetry
at the LHC \cite{Chatrchyan:2012ni}. As we shall shortly argue, multiple branching provides a 
much more efficient mechanism in that sense.
 
 The approximation in Eqs.~(\ref{Ds})--(\ref{E1})  
breaks down when $D^{(1)}(x,\tau)\sim\order{1}$, meaning for $x\lesssim \tau^2$. 
In this non--perturbative regime at small $x$, one needs an exact result which resums 
multiple branching to all orders. Before we present this result,
it is instructive to summarize the picture one would expect
on the basis of our experience with other parton cascades in QCD, like DGLAP. 
(These expectations turn out to be naive, but their failure will be instructive.)

Via successive branchings, the partons at large $x$ get replaced via partons
with smaller values of $x$, which must be numerous enough to carry the energy of their
parents. This seems to imply that the rise in the
gluon distribution $D(x,\tau)$ at small $x$ must become steeper and steeper with increasing  
$\tau$, in such a way to accommodate the energy which disappears at larger values of $x$.
However, this last conclusion is based on the tacit assumption that `the energy remains
in the spectrum', meaning that the energy sum--rule $\int_0^1 \rmd x D(x,\tau)=1$ is satisfied 
at any $\tau$. If that was the case, this would also impose a strong limitation on the energy
that can be carried by the small--$x$ gluons for a given value of $\tau$~: 
in order for the function $D(x,\tau)$ to be integrable as $x\to 0$,
the integral $ {\cal E}(x_0,\tau)\equiv \int_0^{x_0} \rmd x D(x,\tau)$ must vanish as a power
of $x_0$ when $x_0\to 0$, meaning that the energy fraction radiated at large angles could not
be too large. Such a scenario would have little chance to explain the LHC data for di--jet asymmetry. 
However, this is {\em not} the picture that 
emerges from the in--medium evolution equation and that we now describe. 

\begin{figure}[h]
	\centering
	\includegraphics[width=0.6\linewidth]{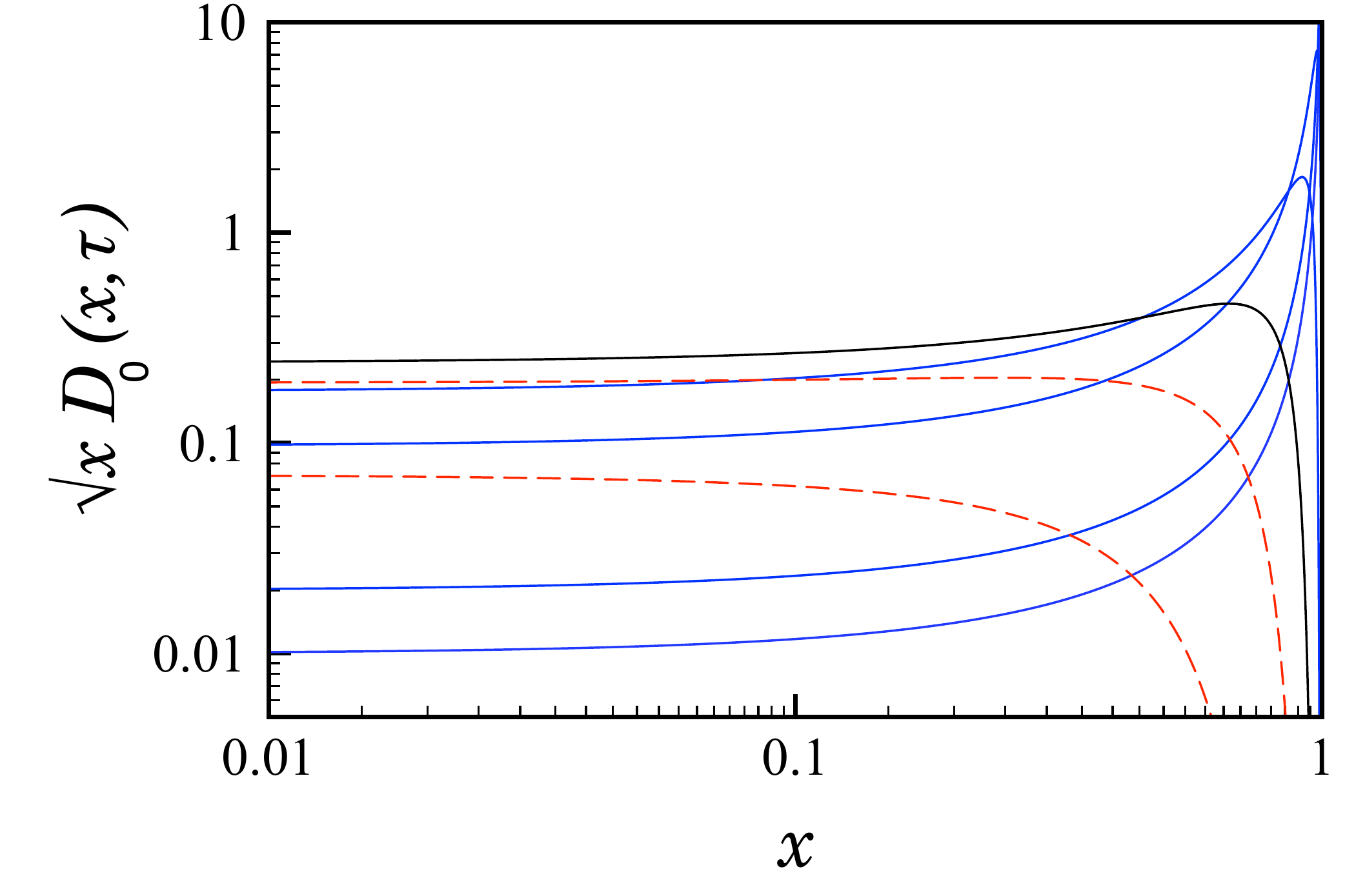}
		\caption{Plot (in Log-Log scale) of $\sqrt{x} D(x,\tau)$ as a function of $x$ for various values of $\tau$ 
		(full lines from bottom to top: $\tau=0.01,0.02,0.1,0.2,0.4$; dashed lines from top down: $\tau=0.6,0.9$).
		}
		\label{fig:spec}

\end{figure}

As mentioned, an exact analytic solution is known, modulo some harmless simplifications in the
branching rate \cite{Blaizot:2013hx}. This is shown here only for $x\ll 1$, where it reads
 \beq\label{Dex}
 D(x\ll 1,\tau)\,\simeq\,\frac{\tau}{\sqrt{x}}\ \rme^{-\pi{\tau^2}}\,.\eeq
(The global spectrum for any $x$ is illustrated in Fig.~\ref{fig:spec} for various values of $\tau$.)
The spectrum (\ref{Dex}) is remarkable in several aspects: \texttt{(i)} The `scaling' behaviour at 
small $x$, $D(x)\propto 1/\sqrt{x}$, is the same as for the BDMPSZ spectrum, \eqn{Ds}.
Formally, one can read \eqn{Dex} as `direct radiation by the leading particle $\times$ survival
probability for the latter'. However, unlike \eqn{Ds}, the spectrum (\ref{Dex}) includes the effects of
multiple branching to all orders. This demonstrates
that the scaling spectrum is a {\em fixed point} of the evolution, for which the `gain' term and the `loss' term
precisely cancel each other. \texttt{(ii)} The fact that the shape of the spectrum at $x\ll 1$ does not change 
with $\tau$ means that its small--$x$ region cannot accommodate the energy which 
disappears via splittings at large $x$. This is also visible in Fig.~\ref{fig:spec}: at small $\tau\ll 1/\sqrt{\pi}$,
the small--$x$ spectrum increases linearly with $\tau$, as the BDMPSZ spectrum (\ref{Ds}).
At the same time, the leading--particle peak, which originally was at $x=1$, moves towards $x<1$ and
gets broader. For larger times $\tau\gtrsim 1/\sqrt{\pi}$, the spectrum is globally suppressed
by the Gaussian factor in (\ref{Dex}). Clearly, the energy disappears from
the spectrum with increasing $\tau$, as confirmed by an explicit calculation of the energy sum--rule \cite{Blaizot:2013hx}: $\int_0^1 \rmd x D(x,\tau)={\rm e}^{-\pi\tau^2}$.  Where does the energy go ?

To answer this question, recall the `fixed--point' property of the scaling spectrum: if one focuses 
on a given bin $x$ with $x\ll 1$, then the amount of energy which enters this bin per unit time due
to splittings at larger $x' > x$ is exactly equal to the amount of energy which leaves that bin
via decays towards smaller $x' < x$. This shows that, via successive branchings, the energy {\em flows 
throughout the entire spectrum without accumulating at any value of $x$}. It therefore accumulates
into a `condensate' ($D_{\rm cond}(x)\propto\delta(x)$) at $x=0$, according to
 \beq\label{Eflow}
 {\cal E}_{\rm flow}(\tau)\,=\,1-\int_0^1 \rmd x D(x,\tau)\,=\, 1- {\rm e}^{-\pi\tau^2}\,.\eeq
So far, we have assumed that the evolution remains unchanged down to $x=0$, but physically
this is not the case: when the gluon energies become
as low as the typical energy scale in the medium --- say, the temperature $T\sim 1$~GeV for
a QGP ---, then the gluons  `thermalize' and disappear from the jet.
Thus, our above conclusion about the `condensate' should be more properly formulated as follows:
{\em via successive branchings, the energy flows towards small--$x$
at a rate which is independent of $x$, and eventually reaches the borderline at $x_{th}\equiv T/E$
between the `jet' and the `medium' at a rate which is independent of the detailed mechanism for 
thermalization and of the precise value of the medium scale $T$.} This energy will be recovered in
the medium at large angles $\theta\gtrsim \theta_{th}$ w.r.t. the jet axis, with $\theta_{th}$ 
obtained by replacing $x_0\sim x_{th}$ within the expression for $\theta_{0}$ shown in \eqn{E1}.

\comment{
{\em Irrespective of the detailed mechanism for thermalization, the energy flows from the jet
towards the medium at a rate which is independent of the thermalization
scale $T$ and thus ends up at small $x\le x_{th}\equiv T/E\ll 1$, meaning at large angles
$\theta\gtrsim \theta_{th}$.} ($\theta_{th}$ can be estimated
by replacing $x_0\to x_{th}$ within the expression of $\theta_{0}$ shown in \eqn{E1}.)}

An energy flow at a rate which is independent of the energy (i.e. uniform in $x$) is the distinguished
signature of {\em wave turbulence} \cite{KST}. This phenomenon is well known in the context
of scalar field theories, but it was not expected in the context of QCD, for the following reason:
as above discussed, the existence of a turbulent flow requires fine cancellations between `gain' and
`loss' terms, which in turn requires the branching process to be {\em quasi--local in $x$}. Or, the
QCD radiation (bremsstrahlung) is reputed for being highly non--local: all the familiar
parton cascades, like DGLAP or BFKL, are dominated by very asymmetric splittings, in which the
splitting fraction $z$ is close to zero, or to one. The medium--induced cascade is new in that respect:
the splitting of a {\em soft} gluon, i.e. the process $x\to zx, (1-z)x$ with $x\ll 1$, is 
controlled by {\em quasi--democratic branchings} with $z\sim 1/2$: the offspring gluons carry
commensurable fractions of the energy of their parent gluon \cite{Blaizot:2013hx}. This is
ultimately related to the peculiar energy dependence of the BDMPSZ spectrum (\ref{spec}),
which in turn reflects the LPM effect.

\begin{figure}[t]
\begin{minipage}{0.35\linewidth}
\centerline{\includegraphics[width=.9\linewidth]{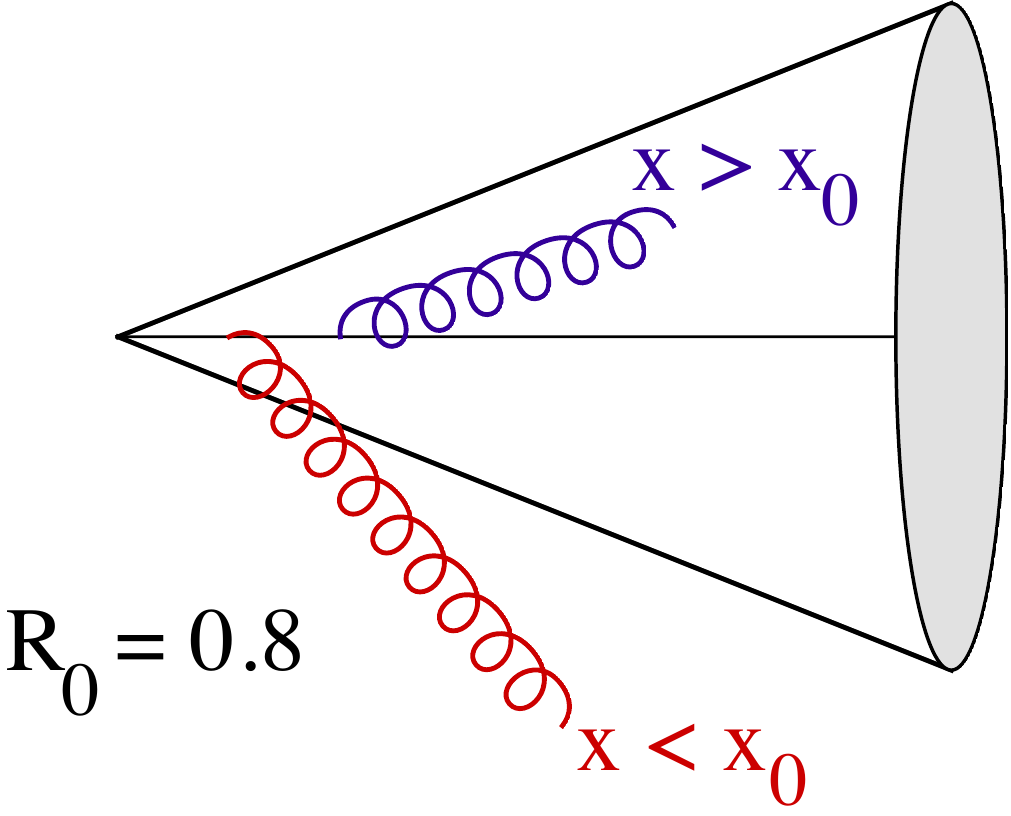}}
\end{minipage}
\hfill
\begin{minipage}{0.65\linewidth}
\centerline{\includegraphics[width=.8\linewidth]{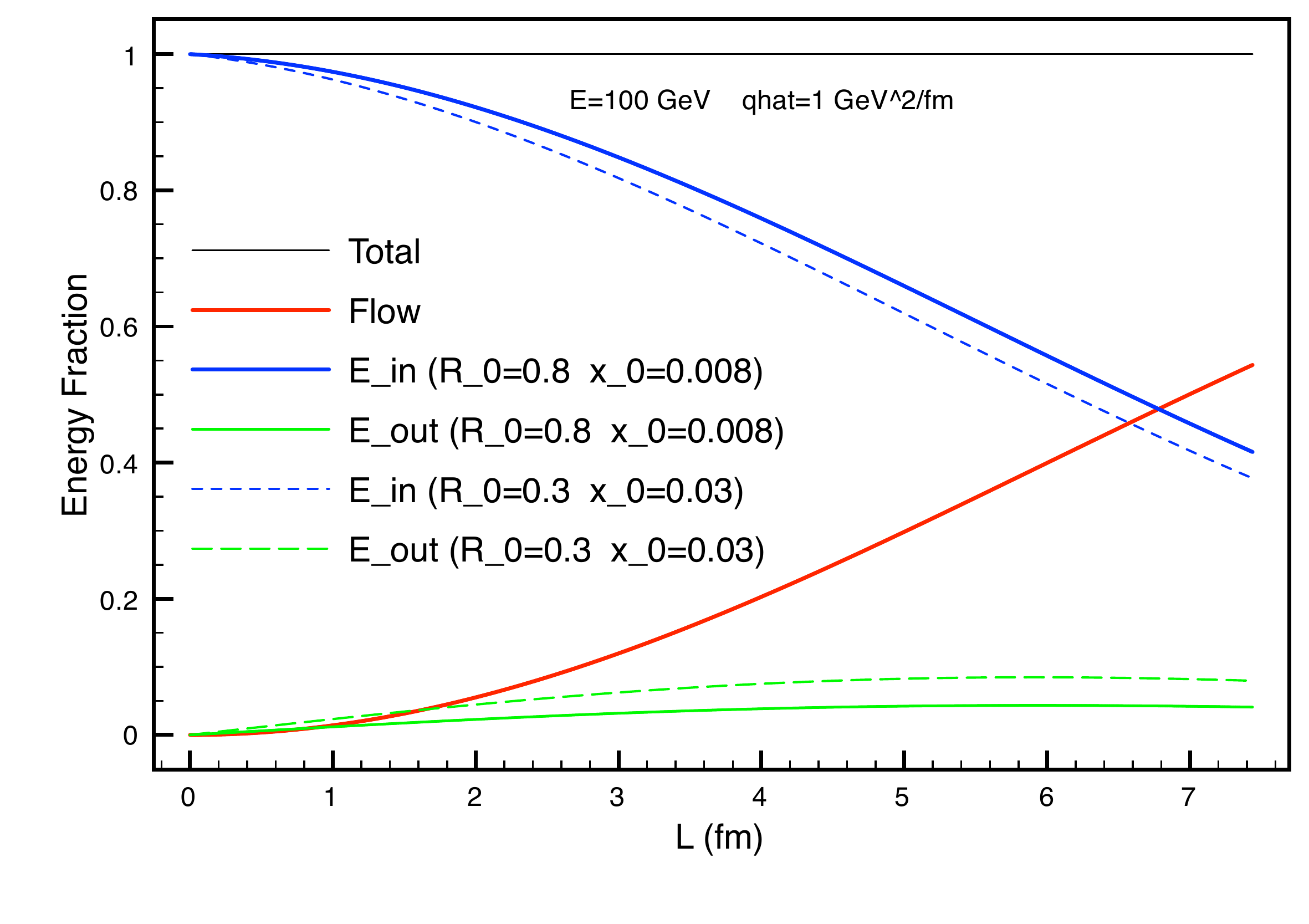}}
\end{minipage}
\caption{The energy balance between the inner part of a conventionaly
defined `jet' with angular opening $R_0$ and the region outside the `jet', plotted 
as a function of the medium size $L$, for two values $R_0=0.3$ and $R_0=0.8$.}
\label{fig:inout}
\end{figure}

The above picture has interesting consequences for the phenomenology.
To see this, let us repeat the exercise in \eqn{E1}, that is, compute the energy radiated at
large angles $\theta > \theta_0$ or, equivalently, small $x< x_0$ (with $x_0\gg x_{th}$ though),
but including the effects of multiple branchings. This includes two contributions: 
the integral over the low--$x$ part of the spectrum (\ref{Dex}) at $x \le x_0$
and the energy (\ref{Eflow}) carried by the turbulent flow. One finds
 \beq\label{Eout}
  {\cal E}(\theta > \theta_0,\tau) \,\simeq\,2\tau \sqrt{x_0}\,{\rm e}^{-\pi\tau^2}\,+\,
  (1- {\rm e}^{-\pi\tau^2})\,\simeq\,2\tau \sqrt{x_0}\,+\,\pi\tau^2\,,
    \eeq
where the second approximation holds for $\pi\tau^2\ll 1$. (Notice that for a jet with
$E=100~{\rm GeV}\approx 2\omega_c$, \eqn{Ds} implies $\tau\simeq\abar\approx 0.3$.)
The flow piece in \eqn{Eout}, which is independent of $x_0$, dominates over the
non--flow piece for any $x_0 < \tau^2$~: {\em the energy lost by the jet at large angles 
$\theta > \theta_0$ is
predominantly carried by the turbulent flow and hence is independent of $\theta_0$}.

This is illustrated in Fig.~\ref{fig:inout} which shows the energy fraction $E_{\rm in}$  {\em inside} a jet
with angular opening $R_0$ ($E_{\rm in}$  is the complement of \eqn{Eout}, 
i.e. the energy contained in the large--$x$ part of the spectrum at $x>x_0$),
together with the two components,  ``Flow'' and ``Non--flow'' (denoted as ``$E_{\rm out}$''),
of the energy fraction {\em outside} the jet, \eqn{Eout}, as functions of the medium size $L$.
As visible there, when increasing the jet angle from $R_0=0.3$ and $R_0=0.8$,
i.e. by almost a factor of 3, the energy captured inside the jet increases only slightly. 
Moreover, $E_{\rm in}$ represents less  than 80\% of the total energy for $L\ge  4$~fm. 
The difference (more than 20\%) is essentially associated with the flow component,
which is independent of $R_0$. This picture is in a remarkable 
agreement with the detailed analysis of the di--jet asymmetry by the CMS
collaboration \cite{Chatrchyan:2012ni}.

\end{document}